\documentclass[aps,prb,twocolumn,reprint,preprintnumbers,amsmath,amssymb]{revtex4-1}
\usepackage{graphicx}% Include figure files
\usepackage{dcolumn}% Align table columns on decimal point
\usepackage{bm}% bold math
\usepackage{amssymb}
\usepackage{epsfig}

\usepackage{epsfig}
\usepackage{amsmath}

\usepackage{amsfonts}
\usepackage{mathptmx}
\usepackage{eucal}

\usepackage{color}
\usepackage[colorlinks,linkcolor=blue,citecolor=blue]{hyperref}
\usepackage{epstopdf}

\usepackage{natbib}
\begin{document}
%\bibliographystyle{prsty}%{apsrev}

%%%%%%%%%%%%%%%%%%%%%%%%%%%%%%%%%%%%%%%%%%%%%%%%%%%%%%%%%%%%
%%%%%%%%%%%%%%%%%%%%%%%%%%%%%%%%%%%%%%%%%%%%%%%%%%%%%%%%%%%%
%%%%%%%%%%%%%%%%%%%%%%%%%%%%%%%%%%%%%%%%%%%%%%%%%%%%%%%%%%%%
%%%%%%%%%%%%%%%%%%%%%%%%%%%%%%%%%%%%%%%%%%%%%%%%%%%%%%%%%%%%
%%%%%%%%%%%%%%%%%%%%%%%%%%%%%%%%%%%%%%%%%%%%%%%%%%%%%%%%%%%%

\title{Realization of quantum diffraction of a thermal flux}

\author{M. J. Mart\'{\i}nez-P\'erez}
\email{mariajose.martinez@sns.it}
\affiliation{NEST, Istituto Nanoscienze-CNR and Scuola Normale Superiore, I-56127 Pisa, Italy}

\author{F. Giazotto}
\email{giazotto@sns.it}
\affiliation{NEST, Istituto Nanoscienze-CNR and Scuola Normale Superiore, I-56127 Pisa, Italy}
%%%%%%%%%%%%%%%%%%%%%%%%%%%%%%%%%%%%%%%%%%%%%%%%%%%%%%%%


%%%%%%%%%%%%%%%%%%%%%%%%%%%%%%%%%%%%%%%%%%%%%%%%%%%%%%%%%%%%
%%%%%%%%%%%%%%%%%%%%%   ABSTRACT         %%%%%%%%%%%%%%%%%%%
%%%%%%%%%%%%%%%%%%%%%   ABSTRACT         %%%%%%%%%%%%%%%%%%%
%%%%%%%%%%%%%%%%%%%%%   ABSTRACT         %%%%%%%%%%%%%%%%%%%
%%%%%%%%%%%%%%%%%%%%%   ABSTRACT         %%%%%%%%%%%%%%%%%%%
%%%%%%%%%%%%%%%%%%%%%   ABSTRACT         %%%%%%%%%%%%%%%%%%%
%%%%%%%%%%%%%%%%%%%%%%%%%%%%%%%%%%%%%%%%%%%%%%%%%%%%%%%%%%%%

\begin{abstract}

\end{abstract}


\maketitle

\textbf{
The first evidence of the dc Josephson\cite{Josephson} effect dates back to 1963 when J. S. Rowell measured the diffraction pattern of the critical current flowing through a single superconducting tunnel junction subjected to an in-plane magnetic field.\cite{rowell}  Interference of Josephson currents through two tunnel junctions connected in parallel was achieved one year later leading to the first ever superconducting quantum interferometer.\cite{Jaklevic} The latter, together with Rowell's observations, constituted the unequivocal demonstration of the Josephson supercurrent-phase relation. Yet, the Josephson effect has further profound implications going beyond electrical transport, as the interplay between the Cooper condensate and unpaired electrons provides thermal flow through the junction with phase coherence as well.\cite{MakiGriffin,GiazottoAPL12,Giazottoarxiv,simmonds} Here we report the first demonstration of quantum diffraction of a heat flux showing that a temperature-biased single Josephson junction is exploited as a diffractor for thermal currents.\cite{GiazottoPRB13}  Specifically, thermal diffraction manifests itself with a peculiar modulation of the electron temperature in a small metallic electrode nearby-contacted to the junction when sweeping the magnetic flux $\Phi$. Remarkably, the observed temperature dependence exhibits  $\Phi$-symmetry and a clear reminiscence with a Fraunhofer-like modulation pattern, as expected fingerprints for a quantum diffraction phenomenon. Our results confirm a pristine prediction of quantum heat transport\cite{GiazottoPRB13}  and, joined with double-junction heat interferometry demonstrated in Ref. \onlinecite{Giazottoarxiv}, exemplify the complementary and conclusive proof of the existence of phase-dependent thermal currents in Josephson-coupled superconductors. Besides shading light on fundamental energy-related aspects in quantum mechanics, this approach combined with well-known methods for phase-biasing superconducting circuits provides with a novel tool for mastering heat fluxes at the nanoscale.\cite{martinezAPL1,martinezAPL2}}

%%%%%%%%%%%%%%%%%%%%%%%%%%%%%%%%%%%%%%%%%%%%%%%%%%%%%%%%%%%
%%%%%%%%%%%%%%%%%%%%   ARTICLE   %%%%%%%%%%%%%%%%%%%%%%%%%%
%%%%%%%%%%%%%%%%%%%%   ARTICLE   %%%%%%%%%%%%%%%%%%%%%%%%%%
%%%%%%%%%%%%%%%%%%%%   ARTICLE   %%%%%%%%%%%%%%%%%%%%%%%%%%
%%%%%%%%%%%%%%%%%%%%   ARTICLE   %%%%%%%%%%%%%%%%%%%%%%%%%%
%%%%%%%%%%%%%%%%%%%%   ARTICLE   %%%%%%%%%%%%%%%%%%%%%%%%%%
%%%%%%%%%%%%%%%%%%%%%%%%%%%%%%%%%%%%%%%%%%%%%%%%%%%%%%%%%%%

\begin{figure}[t]
%\hspace*{-2.em}
\includegraphics[width=0.85\columnwidth]{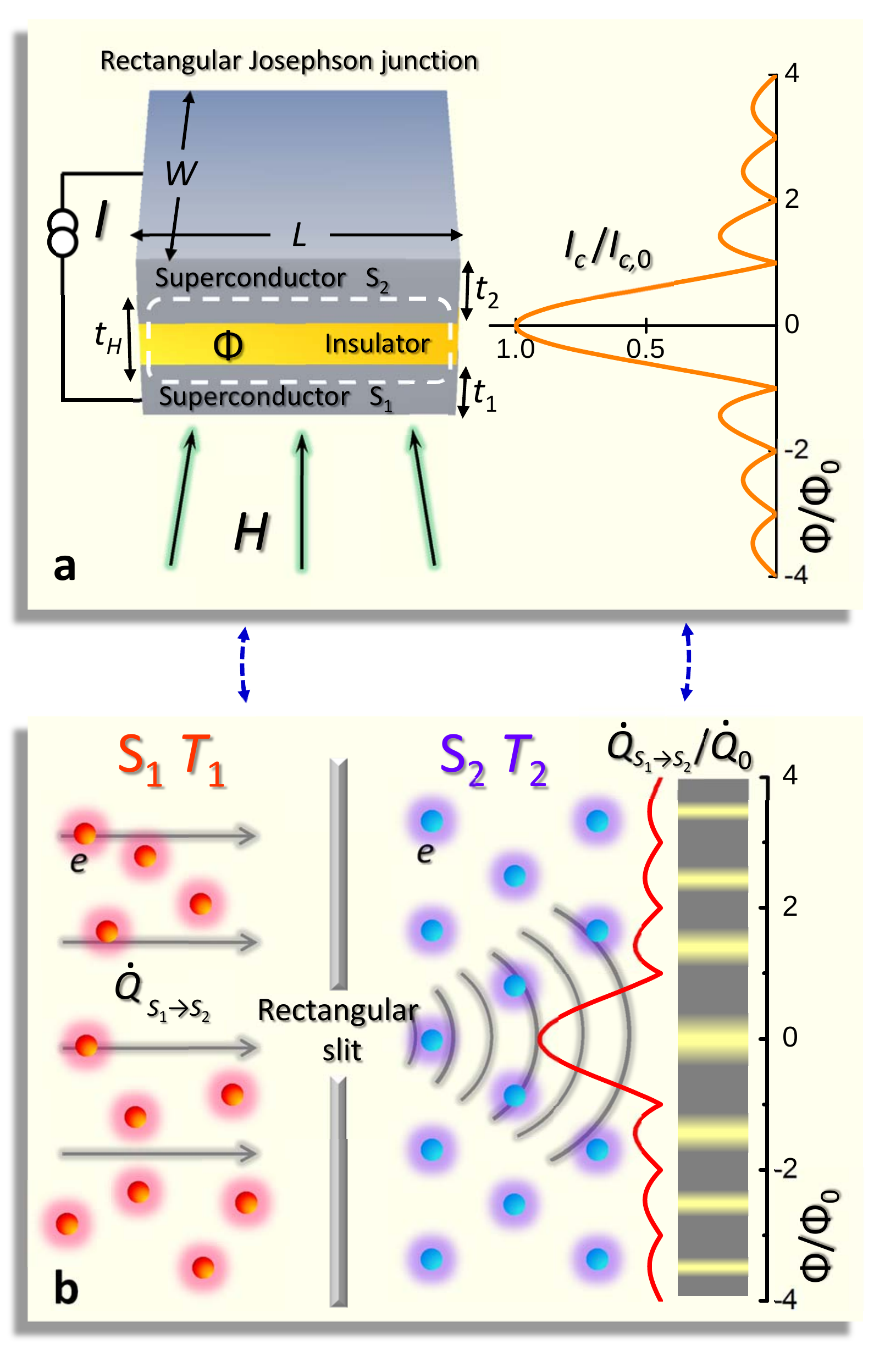}
%\vspace*{-4.ex}
\caption{\textbf{Electric vs. thermal quantum diffraction through a rectangular slit.} (a) The amplitude $I_c$ of the critical current  flowing through a rectangular Josephson junction composed of two superconductors, $S_1$ and $S_2$, separated by a thin insulating layer displays the archetypal Fraunhofer interference pattern as the magnetic flux $\Phi$ threading the junction is varied under an  in-plane sweeping magnetic field $H$. (b) Analogously, when  the two superconductors are kept at different temperatures, $T_{1} > T_{2}$, the resulting heat current $\dot{Q}_{\textrm{S}_1\rightarrow\textrm{S}_2}$ flowing through the junction shows fingerprints of phase coherence. This is reflected, similarly, in a Fraunhofer-like modulation of $\dot{Q}_{\textrm{S}_1\rightarrow \textrm{S}_2}$ with $\Phi$. Both phenomena occur in full analogy to light diffraction through a rectangular slit. $I_{c,0}$ and $\dot{Q}_{0}$ denote the critical and maximum thermal current at zero magnetic field, respectively, whereas $\Phi_0$ is the flux quantum and $I$ the total current flowing through the JJ. $L$, $W$, $t_1$ and $t_2$ denote the junction's main geometrical dimensions and $t_H$ is the effective magnetic thickness defined in the text. }
%\vspace*{-3.ex}
\label{Fig1}
\end{figure} 

\begin{figure*}
%\hspace*{-2.em}
\includegraphics[width=\textwidth]{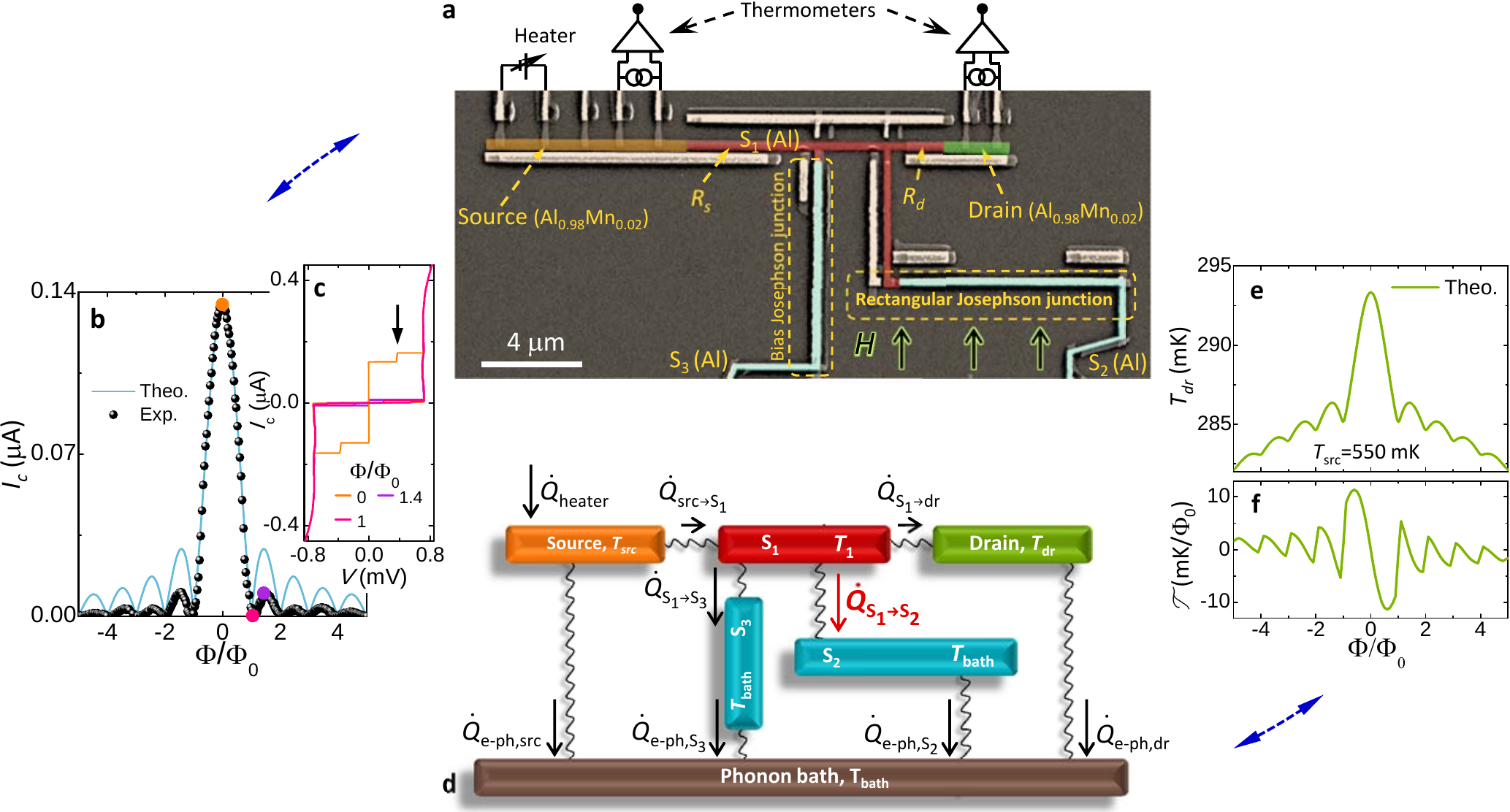}
%\vspace*{-4.ex}
\caption{\textbf{The Josephson thermal quantum diffractor.} (a) Pseudo-color scanning electron micrograph of device A. Thermal diffraction is realized by means of a rectangular Josephson tunnel junction made of two Al superconducting electrodes. The first one ($S_1$) is tunnel-coupled to two source and drain normal metal electrodes (realized with Al$_{0.98}$Mn$_{0.02}$)  enabling Joule heating and thermometry. The second one ($S_2$) extends into a large bonding pad and is kept open during the heat diffraction experiment. The electric characterization of the device is performed through an extra Al probe ($S_3$) tunnel-connected to $S_1$ through a bias JJ. $H$ is the in-plane applied magnetic field. (b) Experimental magnetic diffraction pattern of the critical current $I_c$ (scatter) of the rectangular JJ of device A. Solid line is the theoretical calculation for an ideal rectangular junction. (c) Selected current ($I$) vs. voltage ($V$) curves corresponding to different $\Phi$ values indicated by dots of the same color in panel (b). Curves in (b) and (c) were measured at $240$ mK through the $S_3$-$S_1$-$S_2$ series-connection. (d) Thermal model accounting for the main heat exchange sources present in the device. For clarity, each box is colored as its corresponding electrode in panel (a). Electrons in the source are  intentionally heated up to $T_{\textrm{src}} $ by an injected Joule power,  $\dot{Q}_{ \textrm{heater}}$. Electrons in S$_1$ exchange heat with those in the source at power $\dot{Q}_{\textrm{src}\rightarrow\textrm{S}_1}$, at power $\dot{Q}_{\textrm{S}_1\rightarrow \textrm{S}_2}$ and $\dot{Q}_{\textrm{S}_1\rightarrow\textrm{S}_3}$  with electrons in  S$_2$ and S$_3$, respectively, and at power $\dot{Q}_{ \textrm{S}_1 \rightarrow \textrm{dr}}$ with  electrons in the drain. Finally, electrons in the whole structure exchange energy with lattice phonons residing at $T_{\textrm{bath}}$ at power $\dot{Q}_{\textrm{e-ph},j}$, where $j=$  src, $S_2$, $S_3$ and dr. $S_2$ and $S_3$ are assumed to reside at the bath temperature ($T_{\textrm{bath}}$) owing to their large volume. Arrows indicate the heat flow directions for $T_{\textrm{src}} > T_{1}> T_{\textrm{dr}} > T_{\textrm{bath}}$. (e) Electronic temperature of drain electrode ($T_{\textrm{dr}}$) vs. $\Phi$ calculated using the thermal model described in panel (d) and assuming $T_{\textrm{src}}=550 $ mK and $T_{\textrm{bath}}=240$ mK. Diffraction of thermal currents manifests itself with a peculiar $\Phi$-symmetric function for $T_{dr}$. (f) Flux-to-temperature transfer coefficient,  $ \mathcal{T} = \partial T_{\textrm{dr}} / \partial \Phi$, calculated for the same conditions as in panel (e). }
%\vspace*{-3.ex}
\label{Fig2}
\end{figure*}

\begin{figure*}
%\hspace*{-2.em}
\includegraphics[width=0.7\textwidth]{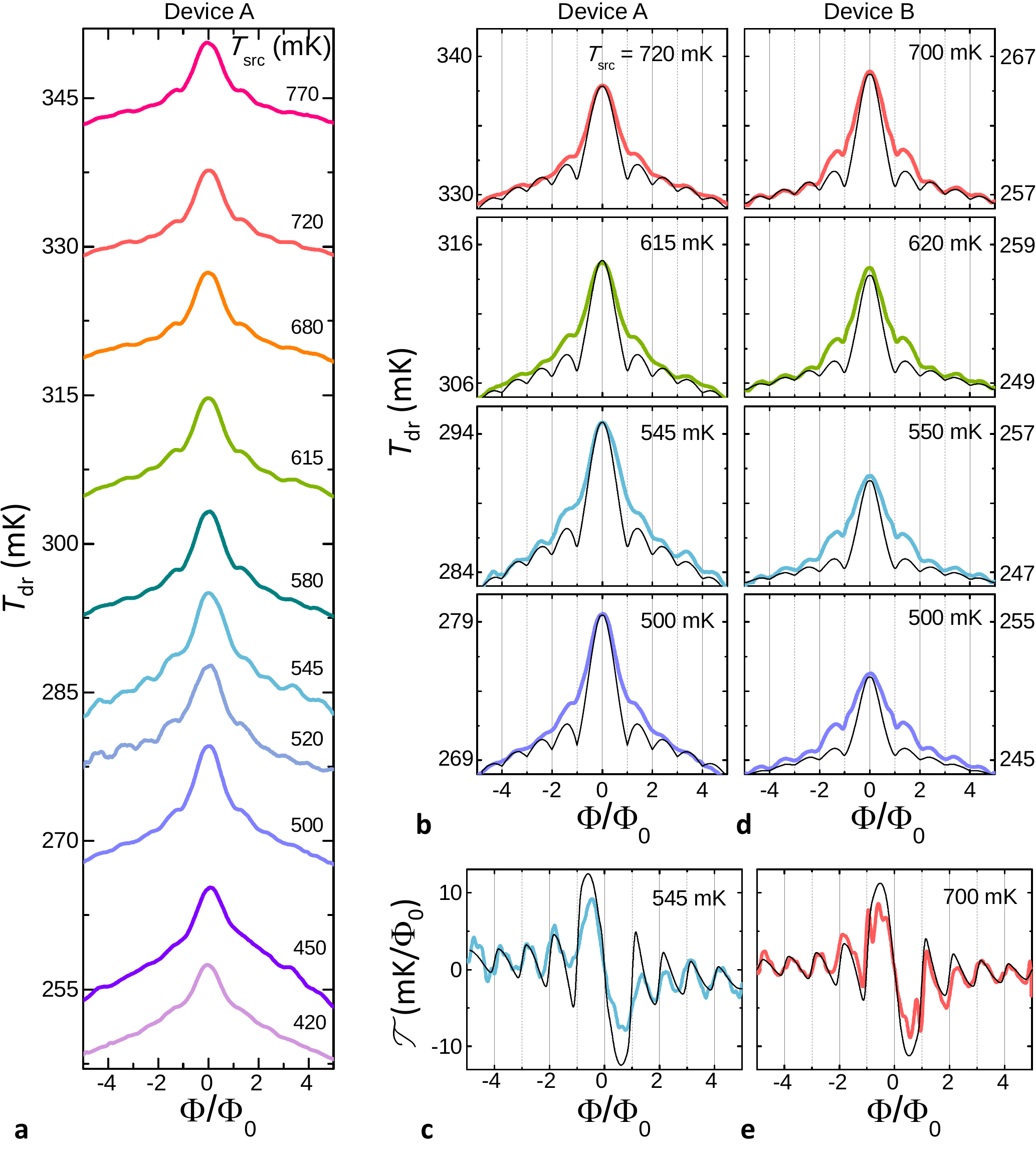}
%\vspace*{-4.ex}
\caption{\textbf{Thermal diffraction at $\textbf{240}$ mK bath temperature.} (a) Gradual increase of $T_{\textrm{dr}}$ vs. $\Phi$ measured at growing $T_{\textrm{src}}$ and $T_{\textrm{bath}} = 240 $ mK for device A. Notably, $T_{\textrm{dr}}$ is $\Phi$-symmetric with a well-defined central lobe surrounded by lumps in the amplitude which decrease as $|\Phi| $ increases, in clear resemblance with a Fraunhofer-like diffraction pattern.  The amplitude of the central lobe increases initially as  $T_{\textrm{src}}$ is raised, decreasing slightly at higher $T_{\textrm{src}}$. Panels (b) and (d) show a few experimental $T_{\textrm{dr}}$ vs. $\Phi$ curves (colour lines) measured at selected values of $T_{\textrm{src}}$ for device A and B, respectively.  The latter is nominally identical in dimensions to sample A and characterized by $R_{\textrm{J}} \approx 580$ $\Omega$, $R_{\textrm{bias}} \approx 480$ $\Omega$, $R_{\textrm{s}} \approx 9.5$ k$\Omega$,  $R_{\textrm{d}} \approx 14$ k$\Omega$ and magnetic flux period $H \approx 37$ Oe.  The vertical scale in each panel is $13$ mK.  Remarkably, $T_{\textrm{dr}}$ exhibits minima at integer multiples of $\Phi_0$ just as the corresponding experimental critical supercurrent diffraction patterns. Black lines are the theoretical curves obtained using the thermal model described in Fig. \ref{Fig2}(d). Panels (c) and (e) display the numerical derivative of the experimental $T_{\textrm{dr}}(\Phi)$ curves at two selected values of $T_{\textrm{src}}$ (coloured lines) and the corresponding calculated flux-to-temperature transfer functions (black lines) for device A and B, respectively.}
%% \vspace*{-3.ex}
\label{Fig3}
\end{figure*}

\begin{figure}[t]
%\hspace*{-2.em}c
\includegraphics[width=0.9\columnwidth]{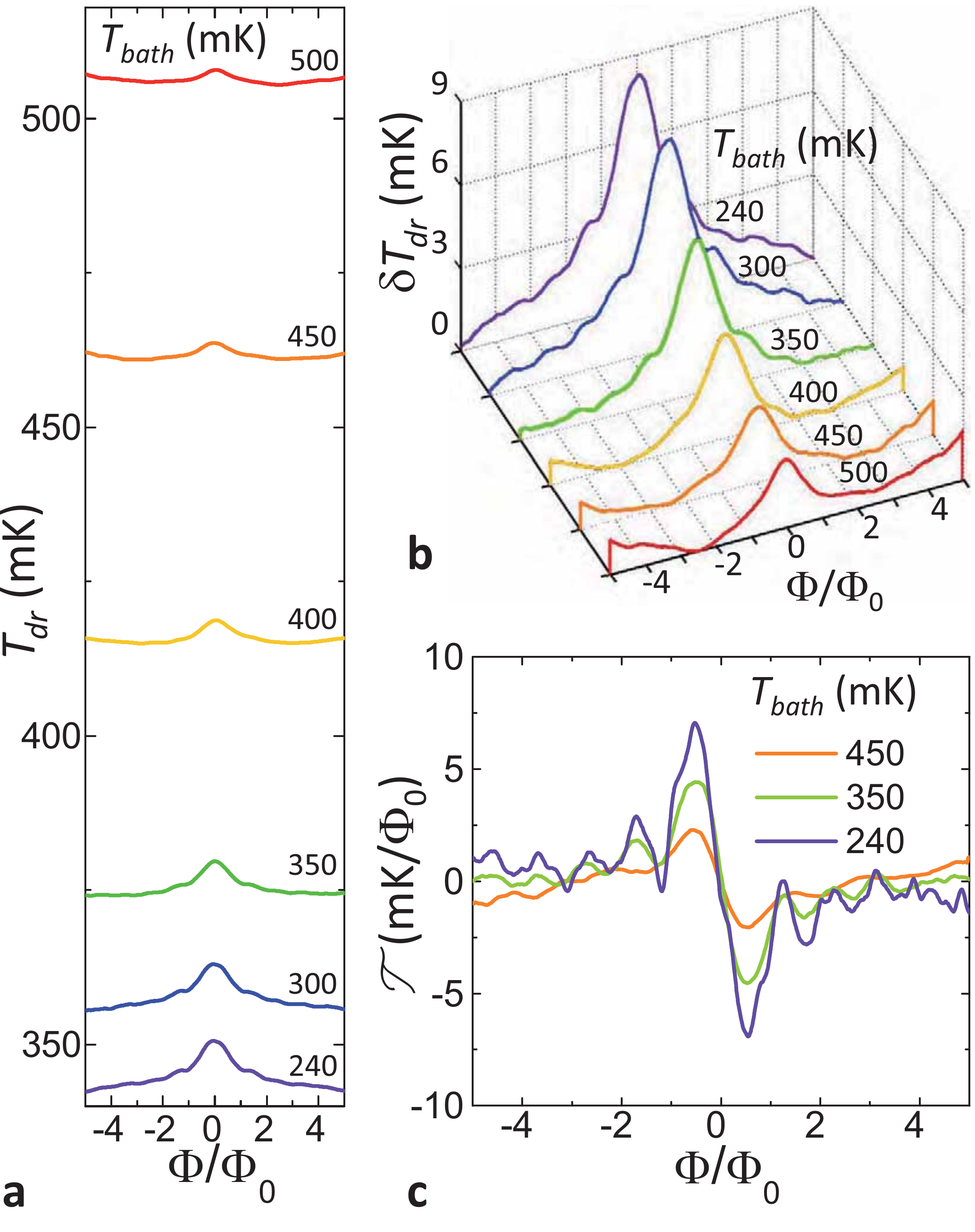}
%\vspace*{-4.ex}
\caption{ \textbf{Thermal diffraction at several bath temperatures.} (a) $T_{\textrm{dr}}$ vs. $\Phi$ characteristics measured at different $T_{\textrm{bath}} $ for device A. From bottom to top, the data were taken at $T_{\textrm{src}}=770$, $780$, $880$, $885$, $920$ and $975$ mK, respectively. These curves are plotted in panel (b) after subtraction of an offset [i.e., $\delta T_{\textrm{dr}} = T_{\textrm{dr}} - \min (T_{\textrm{dr}} )$ ] for each curve to emphasize differences between them. As $T_{\textrm{bath}}$ is raised up, the lobes are clearly smeared while the $\Phi$-symmetry is preserved. $T_{\textrm{dr}}$ modulation fades out for $T_{\textrm{bath}} \gtrsim 500$ mK.  (c)   $\mathcal{T}$ vs. $\Phi$ traces for three selected values of $T_{\textrm{bath}}$.}
%% \vspace*{-3.ex}
\label{Fig4}
\end{figure}

Both electric and thermal quantum diffraction may arise in a solid-state microcircuit by virtue of the Josephson effect. What these phenomena share in common is phase coherence of either supercurrent or thermal flux flowing trough a Josephson junction (JJ). To illustrate this, let us assume an ideal rectangular tunnel JJ composed of two superconductors, $S_1$ and $S_2$, separated by a thin insulating layer under the presence of an in-plane magnetic field $H$. If an electric current $I$ is allowed to flow through the junction, diffraction manifests as the archetypal Fraunhofer interference pattern of the critical current  $I_c$ (see Fig. \ref{Fig1}a).\cite{rowell} By contrast, if the junction is electrically-open but a temperature gradient is applied so that $S_1$ is set at temperature $T_{1}$ while $S_2$ resides at $T_{2}$, a stationary heat current $\dot{Q}_{\textrm{S}_1\rightarrow\textrm{S}_2}$ will develop flowing from  $S_1$ to  $S_2$ (see Fig. \ref{Fig1}b). As predicted in Ref. \onlinecite{GiazottoPRB13} the latter will reflect the consequences of quantum diffraction in full similarity with the electric case. In particular, $\dot{Q}_{\textrm{S}_1\rightarrow\textrm{S}_2}$ is given by\cite{GiazottoPRB13}
\begin{equation}
 \dot{Q}_{\textrm{S}_1\rightarrow \textrm{S}_2} = \dot{Q}_{qp} -\dot{Q}_{int}  \left|  \frac{\sin(\pi \Phi / \Phi_0)}{(\pi \Phi / \Phi_0)}   \right|,
\label{qdottotal}
\end{equation}
where $\Phi_0=2\times10^{-15}$ Wb is the flux quantum. According to Eq. (\ref{qdottotal}), $\dot{Q}_{\textrm{S}_1\rightarrow\textrm{S}_2}$ consists of a Fraunhofer-like diffraction pattern (i.e., the term containing the sine cardinal function) superimposed on top of a magnetic flux-independent heat current. In particular,   $\dot{Q}_{\textrm{S}_1\rightarrow\textrm{S}_2}$ will display minima for integer values of $\Phi_0$ as the critical supercurrent does.
The first term on the rhs of Eq. (\ref{qdottotal}) describes the heat current carried by electrons,\cite{Frank}  $\dot{Q}_{qp}(T_{1},T_{2})=\frac{1}{e^2R_{\textrm{J}}}\int_0^\infty\varepsilon\mathcal{N}_1(\varepsilon,T_{1})\mathcal{N}_2(\varepsilon,T_{2})[f(T_{2})-f(T_{1})]d\varepsilon$, where $R_{\textrm{J}}$ is the normal-state resistance of the JJ, $\mathcal{N}_i(\varepsilon,T_{i})=|\varepsilon|\Theta[\varepsilon^2-\Delta_i(T_{i})^2 ]/ \sqrt{\varepsilon^2-\Delta_i(T_{i})^2}$,\cite{Tinkham}  $f(T_{i})=\tanh(\varepsilon/2k_{\texttt{B}}T_{i})$, $\Delta_i(T_{i})$ is the temperature-dependent energy gap of superconductor $S_i$  with $i=1,2$, $\Theta (x)$ is the Heaviside step function, $k_{\texttt{B}}$ is the Boltzmann constant and $e$ is the electron charge. The second term on the rhs of Eq. (\ref{qdottotal}) is unique to weakly-coupled superconductors and arises from energy-carrying processes involving tunneling of Cooper pairs which leads to its peculiar $\Phi$-dependence. \cite{MakiGriffin,Guttman97,Zhao03,Zhao04} In particular, $\dot{Q}_{int}(T_{1},T_{2})=\frac{1}{e^2R_{\textrm{J}}}\int_0^\infty\varepsilon\mathcal{M}_1(\varepsilon,T_{1})\mathcal{M}_2(\varepsilon,T_{2})[f(T_{2})-f(T_{1})]d\varepsilon$ where  $\mathcal{M}_i(\varepsilon,T_{i})=\Delta_i(T_{i})\Theta[\varepsilon^2-\Delta_i(T_{i})^2]/\sqrt{\varepsilon^2-\Delta_i(T_{i})^2}$. 

A Josephson thermal diffractor (in the following denoted as device A) has been fabricated by electron beam lithography and four-angle shadow mask evaporation of aluminum (Al) and aluminum doped with manganese impurities (Al$_{0.98}$Mn$_{0.02}$). The former constitutes the superconducting electrodes with critical temperature $\approx1.3$ K whereas the latter is a normal metal. The device's core consists of an extended rectangular JJ made of two tunnel-connected Al electrodes, $S_1$ and $S_2$, with $R_{\textrm{J}} \approx 870$ $\Omega$ (see Fig. \ref{Fig2}a). The junction's geometrical dimensions, defined in Fig. \ref{Fig1}a, are $L\approx9$ $\mu$m,  $W\approx0.3$ $\mu$m, $t_1\approx30$ nm and $t_2\approx80$ nm.  $H$ is applied in the junction plane and is perpendicular to its largest lateral dimension, i.e., $L$. An extra aluminum probe $S_3$ is used to current-bias the main JJ for preliminary electric characterization. $S_3$ is connected to $S_1$ through a bias JJ with normal-state resistance $R_{\textrm{bias}}\approx 430$ $\Omega$ placed in orthogonal direction with respect to the main JJ so to be only marginally influenced by $H$. Heat transport through the structure, on the other hand, is investigated thanks to two normal metal source and drain Al$_{0.98}$Mn$_{0.02}$ electrodes  tunnel-connected to $S_1$ wile keeping both JJs \emph{electrically open}. The electronic temperature in the source ($T_{\textrm{src}}$) and in the drain ($T_{\textrm{dr}}$) is experimentally controlled and measured thanks to a number of normal metal-insulator-superconductor (NIS) probes serving as heaters and thermometers.\cite{GiazottoRev,Nahum} Source and drain tunnel junctions have normal-state resistance $R_{\textrm{s}}\approx R_{\textrm{d}}\approx3.5$ k$\Omega$ whereas each NIS probe exhibits $\sim20$ k$\Omega$ on the average.

Quantum diffraction of the \textit{electric} Josephson current is realized first. The resulting experimental $I_c$ vs. $\Phi$ modulation is shown in Fig. \ref{Fig2}b along with the theoretical Fraunhofer diffraction pattern.\cite{Tinkham,barone} $I_c$ is $\Phi$-symmetric   attaining a maximum value of $\approx140$ nA at $\Phi=0$ and nulling at integer values of $\Phi_0$, as expected for a rectangular JJ. Differences in the lobes' amplitude between these curves might reflect non-homogeneous distribution of the supercurrent in the JJ.\cite{barone} These data allow to extract the effective magnetic thickness $t_H$ of the junction defined by the condition  $\Phi=\mu_0HLt_H=\Phi_0$, where $\mu_0$ is the vacuum permeability (see Fig. \ref{Fig1}a). From the experimental magnetic field period $H\approx40$ Oe we get $t_H \approx 57$ nm in good agreement with $59$ nm obtained from geometrical considerations.\cite{Weihnacht} We note that lateral dimensions of the  JJs are much smaller than the Josephson penetration depth, $ \lambda_{\textrm{J}}=\sqrt{\pi \Phi_0LW/\mu_0I_{c,0}t_H}\sim1$ mm, therefore providing the frame of the \textit{short} junction limit.\cite{barone} In such a case, the self-field generated by the Josephson current in the junctions is negligible in comparison to $H$.\cite{GiazottoPRB13} Data in panel \ref{Fig2}b are obtained from the zero-voltage steps in the current ($I$)-voltage  ($V$) characteristics measured through the series connection of the two JJs (see Fig. \ref{Fig2}c). Furthermore, dissipationless electric transport through the main JJ is guaranteed since $R_{\textrm{bias}}<R_{\textrm{J}}$ leading to a larger critical current in the bias JJ.\cite{Tinkham} The ensuing transition of the latter to the dissipative regime  is confirmed by the presence of a second switching step at finite voltage in the $I-V$ characteristics (see black arrow in Fig. \ref{Fig2}c).

On the other hand, quantum diffraction of \emph{thermal} currents is realized as follows. %We assume lattice phonons in the whole structure to reside at the bath temperature $T_{\textrm{bath}}$ so that heat is carried by electrons exclusively.  
A thermal gradient is established by heating intentionally the source's electrons up to a fixed temperature $T_{\textrm{src}}$ leading to an increase on the electronic temperature of  $S_1$ up to $T_1>T_{\textrm{bath}}$. This is possible since $S_1$ is a superconducting electrode with small volume ($\mathcal{V}_{S_1}\approx0.2$ $\mu$m$^3$), allowing for its electrons to be marginally coupled to the lattice phonons at low temperatures.\cite{Timofeev} By contrast, $S_2$ and $S_3$ are strongly thermalized at $T_{\textrm{bath}}$ stemming from their large volume ($\sim10^4$ $\mu$m$^3$).\cite{Timofeev}  Under these circumstances, $T_{\textrm{dr}}$ is mainly determined by the temperature $T_1$ in $S_1$ which is affected by the heat flux $\dot{Q}_{\textrm{S}_1\rightarrow \textrm{S}_2}$. Therefore, $T_{\textrm{dr}}$ can be used to asses the occurrence of thermal diffraction in the main JJ as $H$ is swept. 
 
Insight into this phenomenon can be gained with the help of the thermal model described in Fig. \ref{Fig2}d. $T_{\textrm{ 1}}$ and $T_{\textrm{dr}}$ can be calculated for each $T_{\textrm{src}}$ and $T_{\textrm{bath}}$ fixed in the experiment by solving the following system of two thermal-balance equations (see Methods Summary for further details). The latter accounts for the main heat exchange mechanisms occurring in $S_1$ and drain, respectively; 

\begin{equation}
\begin{array}{lcll}
-\dot{Q}_{\textrm{src}\rightarrow\textrm{S}_1}+\dot{Q}_{\textrm{S}_1\rightarrow \textrm{S}_2}+\dot{Q}_{\textrm{S}_1\rightarrow \textrm{S}_3}+\dot{Q}_{ \textrm{S}_1 \rightarrow \textrm{dr}} =0      \\    
-\dot{Q}_{ \textrm{S}_1\rightarrow\textrm{dr}}+\dot{Q}_{\textrm{e-ph},dr}=0.  
\end{array}
\label{system}
\end{equation}
In writing Eqs. \ref{system} we neglect the electron-phonon heat exchange in $S_1 $ since it is much smaller than that existing in the drain electrode, $\dot{Q}_{\textrm{e-ph},dr}$.\cite{GiazottoRev,Timofeev,Maasiltaa} Heat transport mediated by photons and pure phonon heat current is neglected as well\cite{meschke,schmidt}. As an example, $\Phi$-modulation of $T_{\textrm{dr}}$ is calculated at $T_{\textrm{bath}}=240$ mK using the structure's parameters for $T_{\textrm{src}}=550$ mK. The resulting curve is shown in Fig. \ref{Fig2}e.  Notably, the existence of thermal diffraction leads to a non-monotonic $\Phi$-symmetric function which is maximized at $\Phi=0$ and is suppressed by increasing magnetic flux.  In addition, $T_{\textrm{dr}}(\Phi)$ displays minima exactly at integer values of $\Phi_0$ in close resemblance with a Fraunhofer-like diffraction pattern. Figure \ref{Fig2}f, on the other hand, shows the corresponding magnetic flux-to-temperature transfer coefficient,  $\mathcal{T}=\partial T_{\textrm{dr}}/\partial\Phi$. We stress that the expected temperature modulation arises solely from the combined action of a thermal bias across the JJ and the existence of diffraction of the heat current.  

Thermal diffraction measurements are performed first at the base temperature of a $^3$He refrigerator, i.e., $T_{\textrm{bath}}\approx240$ mK. NIS thermometers in both source and drain electrodes have been calibrated against the cryostat temperature to provide an accurate measure of $T_{\textrm{src}}$ and $T_{\textrm{dr}}$ from the refrigerator base temperature up to $\sim 1$ K. Electron thermometry is performed by current-biasing source and drain SINIS junctions with $70$ and $30$ pA, respectively, so to marginally affect the thermal balance in these electrodes. Source heating, on the other hand, is obtained by delivering a power $\dot{Q}_{\textrm{heater}}$ in the range of $\sim2-100$ pW.

$T_{\textrm{dr}}(\Phi)$ is recorded for different values of $T_{\textrm{src}}$ ranging between $\sim400-800 $ mK and the resulting curves are plotted in Fig. \ref{Fig3}a. The average value of $T_{\textrm{dr}}$ increases as $T_{\textrm{src}}$ is raised up stemming from a larger heat flow induced in the structure. What is more compelling is the peculiar dependence of $T_{\textrm{dr}}$ on $\Phi$ which consists of a sizable peak centered at $\Phi=0$  surrounded by smaller side-lobes preserving $\Phi$-symmetry. These results are  in good resemblance with the theoretical prediction (see Fig. \ref{Fig2}e) therefore pointing to the occurrence of quantum diffraction of the thermal flux. This is further proved in Fig. \ref{Fig3}b where a few selected $T_{\textrm{dr}}(\Phi)$ curves are plotted along with the theoretical expectations (black lines) calculated using the above-described thermal model. Figure \ref{Fig3}c shows the experimental and theoretical flux-to-temperature transfer coefficient at  $T_{\textrm{src}}=545$ mK. Although rather simplified, our model provides a reasonable qualitative agreement with the experiment, and describes the overall $T_{\textrm{dr}}(\Phi)$ modulation shape as well as the exact position of temperature minima. In addition, temperature diffraction measurements have been also performed using a similar sample denoted as device B, leading to comparable results. To illustrate this, Fig. \ref{Fig3}d shows a few selected $T_{\textrm{dr}}(\Phi)$ characteristics along with the corresponding computed ones. The experimental and theoretical $\mathcal{T}(\Phi)$ traces for device B at $T_{\textrm{src}}=700$ mK are plotted in Fig. \ref{Fig3}e. It is worthwhile to recall that the observed thermal diffraction occurs in the absence of any electric current flowing through the JJs. 

The robustness of the $T_{\textrm{dr}}(\Phi)$  modulation against an increasing bath temperature is shown in Fig. \ref{Fig4}a. This leads, on the one hand, to an average enhancement of $T_{\textrm{dr}}$ stemming from a increased total thermal flux through the structure. On the other hand, the amplitude of the modulation decreases and the side-lobes fade out as $T_{\textrm{bath}}$ is raised up. This behavior is emphasized by plotting the $T_{\textrm{dr}}(\Phi)$ curves obtained at different $T_{\textrm{bath}}$ after subtraction of an offset (see Fig. \ref{Fig4}b). A sizable central lobe is still clearly visible also for $T_{\textrm{bath}}>400$ mK but only for considerably higher source temperatures. The same picture is confirmed by inspecting the corresponding  $ \mathcal{T} (\Phi)$ transfer functions (see Fig. \ref{Fig4}c). The visibility of the temperature modulation is somewhat degraded for  $T_{\textrm{bath}}$ exceeding $450$ mK which can be ascribed to both a reduced temperature biasing across the JJs and enhanced electron-phonon coupling in drain electrode at high $T_{\textrm{bath}}$.

Quantum diffraction of a thermal flux has been experimentally realized in a Josephson tunnel junction-based microcircuit. Our results confirm a breaking-new  prediction \cite{GiazottoPRB13} on phase-coherent heat transport and pave the way for the investigation of more exotic junction geometries.\cite{barone,Martucciello,guerlich} These might provide tunable temperature diffraction patterns and should represent a powerful tool for tailoring and managing heat currents at the nanoscale.\cite{GiazottoPRB13,GiazottoAPL08,Ryazanov,Panaitov}  Besides offering insight into energy transport in quantum systems, our experimental findings set the complementary and conclusive demonstration of the ``thermal'' Josephson effect in weakly-coupled superconductors, similarly to what it was done 50 years ago for its ``electric'' counterpart.

%%%%%%%%%%%%%%%%%%%%%%%%%%%%%%%%%%%%%%%%%%%%%%%%%%%%%%%%%%%%
%%%%%%%%%%%%%%%%%%%%   METHODS   %%%%%%%%%%%%%%%%%%%%%%%%%%%
%%%%%%%%%%%%%%%%%%%%   METHODS   %%%%%%%%%%%%%%%%%%%%%%%%%%%
%%%%%%%%%%%%%%%%%%%%   METHODS   %%%%%%%%%%%%%%%%%%%%%%%%%%%
%%%%%%%%%%%%%%%%%%%%   METHODS   %%%%%%%%%%%%%%%%%%%%%%%%%%%
%%%%%%%%%%%%%%%%%%%%   METHODS   %%%%%%%%%%%%%%%%%%%%%%%%%%%
%%%%%%%%%%%%%%%%%%%%%%%%%%%%%%%%%%%%%%%%%%%%%%%%%%%%%%%%%%%%
\vspace{0.5cm}

\textbf{METHODS SUMMARY}
\vspace{0.1cm}

Device A and B are nominally identical and have been fabricated onto an oxidized Si wafer by e-beam lithography of a suspended resist mask and four-angle shadow mask UHV evaporation of metals. The samples are first tilted at $32^\circ$ to deposit a $15$ nm-thick Al$_{0.98}$Mn$_{0.02}$ layer forming source and drain electrodes, and then are exposed to $950$ mTorr of O$_2$ for 5 minutes defining the heater, thermometers, source and drain tunnel barriers. A $20$ nm-thick Al layer is then deposited by tilting the sample at $-49^\circ$ and, subsequently, a second $30$ nm-thick Al layer is evaporated at $32^\circ$ perpendicularly with respect to the previous directions. These two layers define the $S_1$ electrode and the superconducting probes of the NIS junctions. A second oxidation process follows at $1.5$ Torr for 5 minutes to form the JJs tunnel barriers. Finally, a $80$ nm-thick Al layer is evaporated at  $0^\circ$ to define the $S_2$ and $S_3$ electrodes.

Magneto-electric measurements are performed with conventional room-temperature preamplifiers. SINIS thermometers are current biased through battery-powered floating sources whereas the heater operates upon voltage biasing within $0.5-2$ mV. In addition, throughout our measurements we checked that the thermometers response is unaffected by the applied magnetic field. 

In our thermal model [see Eq. (\ref{system})], $\dot{Q}_{\textrm{S}_1\rightarrow \textrm{S}_2}=\alpha\dot{Q}_{qp}-\beta\dot{Q}_{int}\left|\sin(\pi\Phi/\Phi_0)/(\pi\Phi/\Phi_0)\right|$ and $\dot{Q}_{\textrm{S}_1\rightarrow\textrm{S}_3}=\alpha\dot{Q}_{qp}^{\textrm{S}_1\rightarrow\textrm{S}_3}-\beta\dot{Q}_{int}^{\textrm{S}_1\rightarrow\textrm{S}_3}\left|\sin(\pi\Phi_{\textrm{bias}}/\Phi_0)/(\pi\Phi/\Phi_0)\right|$ where $\Phi_{\textrm{bias}}$ denotes the magnetic flux experienced by the bias JJ and  $\alpha$ and $\beta$ are the two fitting parameters. Furthermore $\dot{Q}_{qp}^{\textrm{S}_1\rightarrow\textrm{S}_3}=\frac{1}{e^2R_{\textrm{bias}}}\int_0^\infty\varepsilon\mathcal{N}_1(\varepsilon,T_{1})\mathcal{N}_3(\varepsilon,T_{3})[f(T_{3})-f(T_{1})]d\varepsilon$ and $\dot{Q}_{int}^{\textrm{S}_1\rightarrow\textrm{S}_3}=\frac{1}{e^2R_{\textrm{bias}}}\int_0^\infty\varepsilon\mathcal{M}_1(\varepsilon,T_{1})\mathcal{M}_3(\varepsilon,T_{3})[f(T_{3})-f(T_{1})]d\varepsilon$, with $\mathcal{N}_3(\varepsilon,T_{3})=\mathcal{N}_2(\varepsilon,T_{2})$, $\mathcal{M}_3(\varepsilon,T_{3})=\mathcal{M}_2(\varepsilon,T_{2})$ and $f(T_{3})=f(T_{2})$. On the other hand,  $\dot{Q}_{\textrm{src}(\textrm{S}_1)\rightarrow \textrm{S}_1 (\textrm{dr})}= \frac{1}{e^2 R_{\textrm{s(d)}}}\int_0^\infty\varepsilon\mathcal{N}_1(\varepsilon,T_{1})   [f(T_{1 (\textrm{dr})})-f(T_{\textrm{src}(1)})]d\varepsilon$,   where   $f(T_{\textrm{src(dr)}})=\tanh(\varepsilon/2k_{\texttt{B}}T_{\textrm{src(dr)}})$. Finally, $\dot{Q}_{\textrm{e-ph},dr}=\Sigma\mathcal{V}_{\textrm{dr}}(T_{\textrm{dr}}^6-T_{\textrm{bath}}^6)$ where $\Sigma\approx4\times10^9$ WK$^{-6}$m$^{3}$ is the electron-phonon coupling constant of Al$_{0.98}$Mn$_{0.02}$ experimentally measured for our samples\cite{Maasiltaa,Schmid} and $\mathcal{V}_{\textrm{dr}}\approx2\times10^{-20}$ m$^3$ is drain volume.  To account for the experimental $H$ misalignment, $\Phi_{\textrm{bias}}\sim\Phi/15$ has been used which leads to the peculiar ellipsoidal shape of the $T_ \textrm{dr} (\Phi)$ curves. A quantitative agreement between theory and experiment (see  Fig. \ref{Fig3}) can be achieved only by varying $\alpha$ and $\beta$ between $0.1-1$. The observed deviations might be ascribed to the presence of non idealities in the junctions leading to possible Andreev reflection-dominated heat transport channels,\cite{Zhao04} and to a non-homogeneous  heat current distribution along the structure.

%%%%%%%%%%%%%%%%%%%%%%%%%%%%%%%%%%%%%%%%%%%%%%%%%%%%%%%%%%%%
%%%%%%%%%%%%%%%%%%%%%%   BIBLIOGRAPHY   %%%%%%%%%%%%%%%%%%%%
%%%%%%%%%%%%%%%%%%%%%%   BIBLIOGRAPHY   %%%%%%%%%%%%%%%%%%%%
%%%%%%%%%%%%%%%%%%%%%%   BIBLIOGRAPHY   %%%%%%%%%%%%%%%%%%%%
%%%%%%%%%%%%%%%%%%%%%%   BIBLIOGRAPHY   %%%%%%%%%%%%%%%%%%%%
%%%%%%%%%%%%%%%%%%%%%%   BIBLIOGRAPHY   %%%%%%%%%%%%%%%%%%%%
%%%%%%%%%%%%%%%%%%%%%%%%%%%%%%%%%%%%%%%%%%%%%%%%%%%%%%%%%%%%

%%%%%%%%%%%%%%%%%%%%%%%%%%%%%%%%%%%%%%%%%%%%%%%%%%%%%%%%%%%%
%%%%%%%%%%%%%%%%%%%%   ACKNOWLEDGEMENTS   %%%%%%%%%%%%%%%%%%
%%%%%%%%%%%%%%%%%%%%   ACKNOWLEDGEMENTS   %%%%%%%%%%%%%%%%
%%%%%%%%%%%%%%%%%%%%   ACKNOWLEDGEMENTS   %%%%%%%%%%%%%%%%%%
%%%%%%%%%%%%%%%%%%%%   ACKNOWLEDGEMENTS   %%%%%%%%%%%%%%%%%%
%%%%%%%%%%%%%%%%%%%%   ACKNOWLEDGEMENTS   %%%%%%%%%%%%%%%%%%
%%%%%%%%%%%%%%%%%%%%%%%%%%%%%%%%%%%%%%%%%%%%%%%%%%%%%%%%%%%%
\vspace{0.1cm}

\textbf{Acknowledgements} We acknowledge S. Heun for a careful reading of the manuscript and P. Solinas for comments. The FP7 program No. 228464 "MICROKELVIN", the Italian Ministry of Defense through the PNRM project "TERASUPER", and the Marie Curie Initial Training Action (ITN) Q-NET 264034 are acknowledged for partial financial support.

\vspace{0.1cm}

\textbf{Author Contributions} M. J. M.-P. fabricated the samples, performed the measurements, analyzed the data and carried out the simulations. F. G. conceived the experiment and contributed to the measurements. M. J. M.-P. and F. G. discussed the results and implications at all stages equally, and wrote the paper.

%%%%%%%%%%%%%%%%%%%%%%%%%%%%%%%%%%%%%%%%%%%%%%%%%%%%%%%%%%%%
%%%%%%%%%%%%%%%%%%%%%%%%%%%%%%%%%%%%%%%%%%%%%%%%%%%%%%%%%%%%
%%%%%%%%%%%%%%%%%%%%%%%%%%%%%%%%%%%%%%%%%%%%%%%%%%%%%%%%%%%%
%%%%%%%%%%%%%%%%%%%%%%%%%%%%%%%%%%%%%%%%%%%%%%%%%%%%%%%%%%%%
%%%%%%%%%%%%%%%%%%%%%%%%%%%%%%%%%%%%%%%%%%%%%%%%%%%%%%%%%%%%
%%%%%%%%%%%%%%%%%%%%%%%%%%%%%%%%%%%%%%%%%%%%%%%%%%%%%%%%%%%%
%FIGURES
%%%%%%%%%%%%%%%%%%%%%%%%%%%%%%%%%%%%%%%%%%%%%%%%%%%%%%%%%%%%

\end{document}